\documentclass{article}

\usepackage{arxiv}

\usepackage[utf8]{inputenc} 
\usepackage[T1]{fontenc}    
\usepackage{lmodern}        
\usepackage{hyperref}       
\usepackage{url}            
\usepackage{booktabs}       
\usepackage{amsfonts}       
\usepackage{nicefrac}       
\usepackage{microtype}      
\usepackage{graphicx}

\title{Frame to frame interpolation for high-dimensional data
visualisation using the woylier package}

\author{
    Zoljargal Batsaikhan
   \\
    Department of Econometrics and Business Statistics \\
    Monash University \\
   \\
  \texttt{\href{mailto:zoljargal11@gmail.com}{\nolinkurl{zoljargal11@gmail.com}}} \\
   \And
    Dianne Cook
   \\
    Department of Econometrics and Business Statistics \\
    Monash University \\
   \\
  \texttt{\href{mailto:dicook@monash.edu}{\nolinkurl{dicook@monash.edu}}} \\
   \And
    Ursula Laa
   \\
    Institute of Statistics \\
    University of Natural Resources and Life Sciences Vienna \\
   \\
  \texttt{\href{mailto:ursula.laa@boku.ac.at}{\nolinkurl{ursula.laa@boku.ac.at}}} \\
  }

\usepackage{color}
\usepackage{fancyvrb}

\DefineVerbatimEnvironment{Highlighting}{Verbatim}{commandchars=\\\{\}}
\usepackage{framed}
\definecolor{shadecolor}{RGB}{248,248,248}
\newenvironment{Shaded}{\begin{snugshade}}{\end{snugshade}}

\newcommand{\AttributeTok}[1]{\textcolor[rgb]{0.77,0.63,0.00}{#1}}

\newcommand{\DecValTok}[1]{\textcolor[rgb]{0.00,0.00,0.81}{#1}}

\newcommand{\FunctionTok}[1]{\textcolor[rgb]{0.00,0.00,0.00}{#1}}

\newcommand{\NormalTok}[1]{#1}

\newcommand{\SpecialCharTok}[1]{\textcolor[rgb]{0.00,0.00,0.00}{#1}}

\providecommand{\tightlist}{%
  \setlength{\itemsep}{0pt}\setlength{\parskip}{0pt}}

\newlength{\cslhangindent}
\newlength{\csllabelwidth}
\newlength{\cslentryspacingunit} 
  {}%
  {\par}
\newenvironment{CSLReferences}[2] 
 {
  \setlength{\parindent}{0pt}
  \ifodd #1
  \let\oldpar\par
  \def\par{\hangindent=\cslhangindent\oldpar}
  \fi
  \setlength{\parskip}{#2\cslentryspacingunit}
 }%
 {}
\usepackage{calc}

\usepackage{booktabs}
\usepackage{longtable}
\usepackage{array}
\usepackage{multirow}
\usepackage{wrapfig}
\usepackage{float}
\usepackage{colortbl}
\usepackage{pdflscape}
\usepackage{tabu}
\usepackage{threeparttable}
\usepackage{threeparttablex}
\usepackage[normalem]{ulem}
\usepackage{makecell}
\usepackage{xcolor}
\begin{document}
\maketitle

\begin{abstract}
The woylier package implements tour interpolation paths between frames
using Givens rotations. This provides an alternative to the geodesic
interpolation between planes currently available in the tourr package.
Tours are used to visualise high-dimensional data and models, to detect
clustering, anomalies and non-linear relationships. Frame-to-frame
interpolation can be useful for projection pursuit guided tours when the
index is not rotationally invariant. It also provides a way to
specifically reach a given target frame. We demonstrate the method for
exploring non-linear relationships between currency cross-rates.
\end{abstract}

\hypertarget{introduction}{%
\section{Introduction}\label{introduction}}

When data has up to three variables, visualization is relatively
intuitive, while with more than three variables, we face the challenge
of visualizing high dimensions on 2D displays. This issue was tackled by
the \emph{grand tour} (Asimov 1985) which can be used to view data in
more than three dimensions using linear projections. It is based on the
idea of rotations of a lower-dimensional projection in high-dimensional
space. The grand tour allows users to see dynamic low-dimensional
(typically 2D) projections of higher-dimensional space. Originally,
Asimov's grand tour presented the viewer with an automatic movie of
projections with no user control. Since then new work has added
interactivity to the tour, giving more control to users (Buja et al.
2005). New variations include the manual (Cook and Buja 1997) or radial
tour (Laa et al. 2023), little tour, guided tour (Cook et al. 1995),
local tour, and planned tour. These are different ways of selecting the
sequence of projection bases for the tour, for an overview see Lee et
al. (2022).

The guided tour combines projection pursuit with the grand tour and it
is implemented in the \CRANpkg{tourr} package (Wickham et al. 2011).
Projection pursuit is a procedure used to locate the projection of
high-to-low dimensional space that should expose the most interesting
feature of data, originally proposed in Kruskal (1969). It involves
defining a criterion of interest, a numerical objective function that
indicates the interestingness of each projection, and an optimization
for selecting planes with increasing values of the function. In the
literature, a number of such criteria have been developed based on
clustering, spread, and outliers.

A tour path is a sequence of projections and we use an interpolation to
produce small steps simulating a smooth movement. The current
implementation of tour in the \CRANpkg{tourr} package uses geodesic
interpolation between planes. The geodesic interpolation path is the
shortest path between planes with no within-plane spin (see Buja et al.
(2004) for more details). As a result, the rendered target plane could
be a within-plane rotation of the target plane originally specified.
This is not a problem when the structure we are looking for can be
identified from any rotation. However, even simple associations in 2D,
such as the calculated correlation between variables, can be very
different when the basis is rotated.

Most projection pursuit indexes, particularly those provided by in
\CRANpkg{tourr} are rotationally invariant. However, there are some
applications where the orientation of frames does matter. One example is
the splines index proposed by Grimm (2016). The splines index computes a
spline model for the two variables in a projection (using the
implementation in \CRANpkg{mgcv} (Wood 2011)), in order to measure
non-linear association. It compares the variance of the response
variable to the variance of residuals, and the functional dependence is
stronger when the index value is larger. It can be useful to detect
non-linear relationships in high-dimensional data. However, its value
will change substantially if the projection is rotated within the plane
(Laa and Cook 2020). The procedure in Grimm (2016) was less affected by
the orientation because it considered only pairs of variables, and it
selected the maximum value found when exchanging which variable is
considered as a predictor or response variable.

\begin{figure}
\includegraphics[width=1\linewidth]{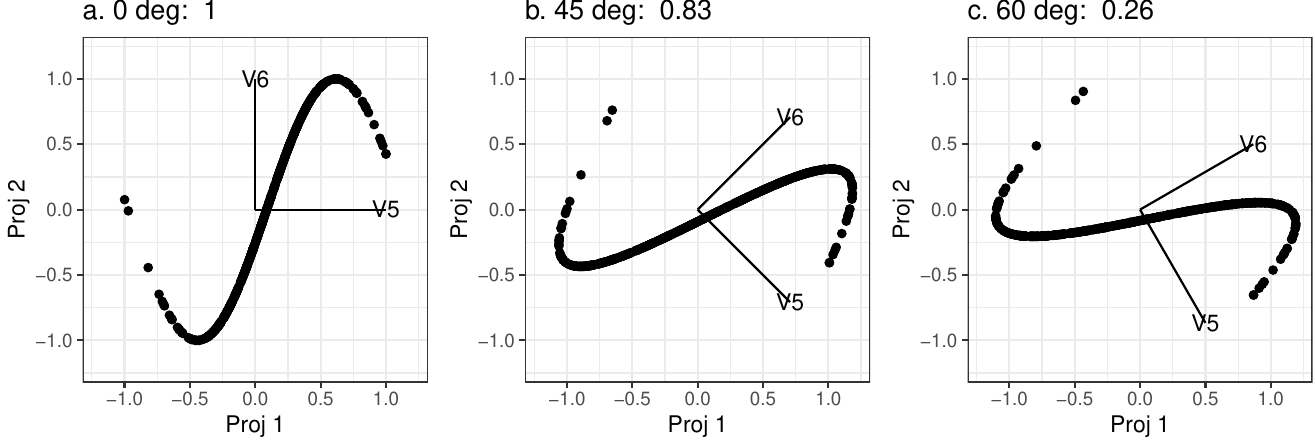} \caption{The impact of rotation on a spline index that is NOT rotation invariant. The index value for different within-plane rotations take very different values: (a) original projection has a maximum index value of 1.00, (b) axes rotated 45$^o$ drops index value to 0.83, (c) axes rotated 60$^o$ drops index to a very low 0.26. Geodesic interpolation between planes will have difficulty finding the maximum of an index like this because it is focused only on the projection plane, not the frame defining the plane.}\label{fig:splines2d-static}
\end{figure}

Figure \ref{fig:splines2d-static} illustrates the rotational invariance
problem for a modified splines index, where we always consider the
horizontal direction as the predictor variable and the vertical
direction as the response. Thus, our modified index computes the splines
on one orientation, exaggerating the rotational variability. The example
data was simulated to follow a sine curve and the modified splines index
is calculated on different within-plane rotations of the data. Although
they have the same structure, the index values vary greatly.

The lack of rotation invariance of the splines index raises
complications in the optimization process in the projection-pursuit
guided tour as available in \CRANpkg{tourr}. Fixing this is the
motivation of this work. The goal with the frame-to-frame interpolation
is that optimization would find the best within-plane rotation, and thus
appropriately optimize the index.

A few alternatives to geodesic interpolation were proposed by Buja et
al. (2005) including the decomposition of orthogonal matrices, Givens
decomposition, and Householder decomposition. The purpose of the
\CRANpkg{woylier} package is to implement the Givens paths method in R.
This algorithm adapts the Given's matrix decomposition technique which
allows the interpolation to be between frames rather than planes.

This article is structured as follows. The next section provides the
theoretical framework of the Givens interpolation method followed by a
section about the implementation in R. The method is applied to search
for non-linear associations between currency cross-rates.

\hypertarget{background}{%
\section{Background}\label{background}}

The tour method of visualization shows a movie that is an animated
high-to-low dimensional data rotation. It is a one-parameter (time)
family of static projections. Algorithms for such dynamic projections
are based on the idea of smoothly interpolating a discrete sequence of
projections (Buja et al. 2005).

The topic of this article is the construction of the paths of
projections. The interpolation of these paths can be compared to
connecting line segments that interpolate points in Euclidean space.
Interpolation acts as a bridge between a continuous animation and the
discrete choice of sequences of projections.

\textbf{Interpolating paths of planes versus paths of frames}

\begin{figure}

{\centering \includegraphics[width=1\linewidth]{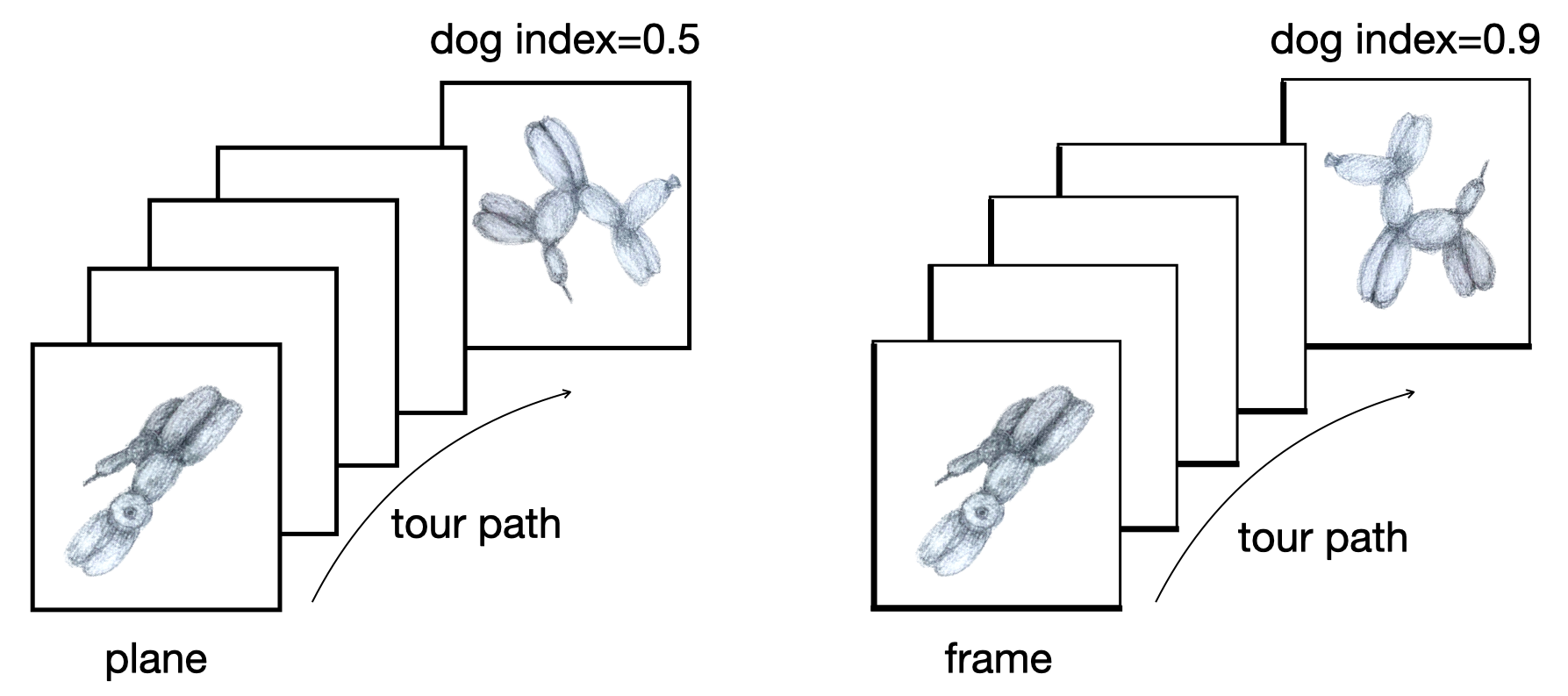} 

}

\caption{Plane to plane interpolation (left) and frame to frame interpolation (right). We use the dog index for illustration purposes. Orientation of the data within the plane could affect the index value. Frame to frame interpolation guarantees reaching a particular frame of the many defining a plane.}\label{fig:dogs}
\end{figure}

The \CRANpkg{tourr} package implements geodesic interpolation between
planes, and the final interpolation step will reach the rotation of the
target frame, avoiding any within-plane spin along the path. When the
the orientation of projections matters interpolation between frames is
required. The orientation of the frames could be important when a
non-linear projection pursuit index function is used in the guided tour.
This is illustrated by the different index values shown in the sketch in
Figure \ref{fig:dogs}, as well as the splines index for the sine curve
in Figure \ref{fig:splines2d-static}.

To describe the interpolation algorithms we will use the following
notation.

\begin{itemize}
\item
  Let \(p\) be the dimension of original data and \(d\) be the dimension
  onto which the data is being projected.
\item
  A frame \(F\) is defined as a \(p\times d\) matrix with pairwise
  orthogonal columns of unit length that satisfies \[F^TF = I_d,\] where
  \(I_d\) is the identity matrix in d dimensions.
\item
  Paths of frames are given by continuous one-parameter families
  \(F(t)\) where \(t\in [a, z]\) represents time. We denote the starting
  frame (at time \(a\)) by \(F_a = F(a)\) and target frame (at time
  \(z\)) by \(F_z = F(z)\). Usually, \(F_z\) is the target frame that
  has been chosen according to the selected tour method. While a grand
  tour chooses target frames randomly, the guided tour chooses the
  target frame by optimizing the projection pursuit index. Interpolation
  methods are used to find the path that moves from \(F_a\) to \(F_z\).
\end{itemize}

\textbf{Preprojection algorithm}

In order to make the interpolation algorithm simple, we carry out a
preprojection step to find the subspace that the interpolation path,
\(F(t)\), is traversing. In other words, the preprojection step defines
the joint subspace of \(F_a\) and \(F_z\) and makes sure the
interpolation path is limited to that space.

The procedure starts with forming an orthonormal basis by applying
Gram-Schmidt to \(F_z\) with regards to \(F_a\), i.e.~we find the
\(p\times d\) matrix that contains the component of \(F_z\) that is
orthogonal to \(F_a\). We denote this orthonormal basis by \(F_\star\).
Then we build the preprojection basis \(B\) by combining \(F_a\) and
\(F_\star\) as follows:

\[B = (F_a, F_{\star})\]

The dimension of the resulting orthonormal basis, \(B\), is
\(p\times 2d\).

Then, we can express the original frames in terms of this basis:

\[F_a = B W_a, F_z = B W_z\]

The interpolation problem is then reduced to the construction of paths
of frames \(W(t)\) that interpolates between the preprojected frames
\(W_a\) and \(W_z\). By construction, \(W_a\) is a \(2d\times d\) matrix
of 1s and 0s. This is an important characteristic of our interpolation
algorithm of choice, the Givens interpolation.

\textbf{Givens interpolation path algorithm}

A rotation matrix is a transformation matrix used to perform a rotation
in Euclidean space. The matrix that rotates a 2D plane by an angle
\(\theta\) looks like this:

\[ \begin{bmatrix}\cos \theta &-\sin \theta \\\sin \theta &\cos \theta \end{bmatrix} \]

If the rotation is in the plane of two selected variables, it is called
a Givens rotation. Let's denote those 2 variables as \(i\) and \(j\).
The Givens rotation is used for introducing zeros, for example when
computing the QR decomposition of a matrix in linear algebra problems.

The interpolation method in the \CRANpkg{woylier} package is based on
the fact that in any vector of a matrix, one can zero out the \(i\)-th
coordinate with a Givens rotation in the \((i, j)\)-plane for any
\(j\neq i\) (Golub and Loan 1989). This rotation affects only
coordinates \(i\) and \(j\) and leaves all other coordinates unchanged.
Sequences of Givens rotations can map any orthonormal d-frame \(F\) in
p-space to the standard d-frame
\[E_d=((1, 0, 0, ...)^T, (0, 1, 0, ...)^T, ...).\]

The resulting interpolation path construction algorithm from starting
frame \(F_a\) to target frame \(F_z\) is illustrated below. The example
is for \(p=6\) and \(d=2\).

\begin{enumerate}
\def\labelenumi{\arabic{enumi}.}
\tightlist
\item
  Construct preprojection basis \(B\) by orthonormalizing \(F_z\) with
  regards tp \(F_a\) with Gram-Schmidt.
\end{enumerate}

In our example, \(F_a\) and \(F_z\) are \(p\times d\) or \(6\times2\)
matrices that are orthonormal. The preprojection basis \(B\) is
\(p\times 2d\) matrix that is \(6\times 4\).

\begin{enumerate}
\def\labelenumi{\arabic{enumi}.}
\setcounter{enumi}{1}
\tightlist
\item
  Get the preprojected frames using the preprojection basis \(B\).
  \[W_a = B^TF_a = E_d\] and \[W_z = B^TF_z\]
\end{enumerate}

In our example, \(W_a\) looks like:

\[ \begin{bmatrix}1 & 0 \\0  &1 \\ 0&0 \\0&0\end{bmatrix} \]

\(W_z\) is an orthonormal \(2d\times d\) matrix that looks like:

\[ \begin{bmatrix} a_{11} & a_{12} \\a_{21}  &a_{22} \\ a_{31}&a_{32} \\a_{41}&a_{42}\end{bmatrix} \]

\begin{enumerate}
\def\labelenumi{\arabic{enumi}.}
\setcounter{enumi}{2}
\tightlist
\item
  Then, we can construct a sequence of Givens rotations that maps
  \(W_z\) to \(W_a\) with such angles that makes one element zero at a
  time:
\end{enumerate}

\[ W_a = R_m(\theta_m) ... R_2(\theta_2)R_1(\theta_1)W_z\]

At each rotation, the angle \(\theta_i\) that zeros out the next
coordinate of a plane is calculated. Here \(m = \sum_{k=1}^d (2d - k)\),
so when \(d=2\) we need \(m=5\) rotations with 5 different angles, each
making one element 0. For example, the first rotation angle \(\theta_1\)
is an angle in radians between \((1, 0)\) and \((a_{11}, a_{21})\). This
rotation matrix would make element \(a_{21}\) zero:

\[R_1(\theta_1) = G(1, 2, \theta_1) = \begin{bmatrix} cos\theta_1 & -sin\theta_1 & 0 & 0 \\sin\theta_1  &cos\theta_1 & 0 &0 \\ 0&0&1&0 \\0&0&0&1\end{bmatrix}\]
Here \(G(i,j,\theta_k)\) denotes a Givens rotation in components \(i\)
and \(j\) by angle \(\theta_k\). In the same way, we zero out the
elements \(a_{31}\) and \(a_{41}\). Because of the orthonormality this
means that now \(a_{11} = 1\) and that \(a_{12} = 0\). We thus need only
two more rotations to zero out \(a_{32}\) and \(a_{42}\).

Each \(\theta_i\) is an angle in 2D, and is computed from the polar
coordinates returned by the \texttt{atan2()} function.

\begin{enumerate}
\def\labelenumi{\arabic{enumi}.}
\setcounter{enumi}{3}
\tightlist
\item
  The inverse mapping is obtained by reversing the sequence of rotations
  with the negative of the angles, we start from the starting basis and
  end at the target basis.
\end{enumerate}

\[R(\theta) = R_1(-\theta_1) ... R_m(-\theta_m), \    W_z = R(\theta)W_a\]

Performing these rotations would go from the starting frame to the
target frame in one step. But we want to do it sequentially in a number
of steps so interpolation between frames looks dynamic.

\begin{enumerate}
\def\labelenumi{\arabic{enumi}.}
\setcounter{enumi}{4}
\item
  Next we include the time parameter, \(t\), so that the interpolation
  process can be rendered in the movie-like sequence. We break each
  \(\theta_k\) into the number of steps, \(n_{step}\), that we want to
  take from the starting frame to the target frame, which means it moves
  by equal angle in each step. Here \(n_{step}\) should vary based on
  the angular distance between \(F_a\) and \(F_z\), such that when
  watching a sequence of interpolations we have a fixed angular speed.
\item
  Finally, we reconstruct our original frames using \(B\). This
  reconstruction is done at each step of interpolation so that we have
  the interpolated path of frames as the result.
\end{enumerate}

\[F_t = B  W_t\]

At each time \(t\) we can project the data using the frame \(F_t\).

\hypertarget{implementation}{%
\section{Implementation}\label{implementation}}

We implemented each of the steps in the Givens interpolation path
algorithm in separate functions and combined them into a single function
\texttt{givens\_full\_path()} to produce the full set of \(F_t\). The
same functions are used to integrate the Givens interpolation with the
\texttt{animate()} functions of the \CRANpkg{tourr} package. Table
\ref{tab:fns-pd} lists the input and output of each function and its
descriptions, functions to use with \texttt{animate()} are described
separately below.

\begin{table}

\caption{\label{tab:fns-pdf}Primary functions in the woylier package.}
\centering
\begin{tabular}[t]{>{\raggedright\arraybackslash}p{5cm}|>{\raggedright\arraybackslash}p{3cm}|>{\raggedright\arraybackslash}p{2cm}|>{\raggedright\arraybackslash}p{2cm}}
\hline
\textbf{name} & \textbf{description} & \textbf{input} & \textbf{output}\\
\hline
\ttfamily{givens\_full\_path(Fa, Fz, nsteps)} & Construct full interpolated frames. & Starting and target frame (Fa, Fz) and number of steps & An array with nsteps matrix. Each matrix is interpolated frame in between starting and target frames.\\
\hline
\ttfamily{preprojection(Fa, Fz)} & Build a d-dimensional pre-projection space by orthonormalizing Fz with regard to Fa. & Starting and target frame (Fa, Fz) & B pre-projection p x 2D matrix\\
\hline
\ttfamily{construct\_preframe(Fa, B)} & Construct preprojected frames. & A frame and the pre-projection p x 2D matrix & Pre-projected frame in pre-projection space\\
\hline
\ttfamily{row\_rot(a, i, k, theta)} & Performs Givens rotation . & A frame and the pre-projection p x 2D matrix & theta angle rotated matrix a\\
\hline
\ttfamily{calculate\_angles(Wa, Wz)} & Calculate angles of required rotations to map Wz to Wa. & Preprojected frames (Wa, Wz) & Names list of angles\\
\hline
\ttfamily{construct\_moving\_frame(Wt, B)} & Reconstruct interpolated frames using pre-projection. & Pre-projection matrix B, Each frame of givens path & A frame of on a step of interpolation\\
\hline
\end{tabular}
\end{table}

When using \CRANpkg{tourr} we typically want to run a tour live, such
that target selection and interpolation are interleaved, and the display
will show the data for each frame \(F_t\) in the interpolation path. The
implementation in \CRANpkg{tourr} was described in Wickham et al.
(2011), and with \CRANpkg{woylier} we provide functions to use the
Givens interpolation with the grand tour, guided tour and planned tour.
To do this we rely primarily on the function \texttt{givens\_info()}
which calls the functions listed in the Table above and collects all
necessary information for interpolating between a given starting and
target frame. The function \texttt{givens\_path()} then defines the
interpolation and can be used instead of
\texttt{tourr::geodesic\_path()}. Wrapper functions for the different
tour types are available to use this interpolation, since in the
\texttt{tourr::grand\_tour()} and other path functions this is fixed to
use the geodesic interpolation. Calling a grand tour with Givens
interpolation for direct animation will then use:

\begin{Shaded}
\begin{Highlighting}[]
\NormalTok{tourr}\SpecialCharTok{::}\FunctionTok{animate\_xy}\NormalTok{(}\SpecialCharTok{\textless{}}\NormalTok{data}\SpecialCharTok{\textgreater{}}\NormalTok{, }\AttributeTok{tour\_path =}\NormalTok{ woylier}\SpecialCharTok{::}\FunctionTok{grand\_tour\_givens}\NormalTok{())}
\end{Highlighting}
\end{Shaded}

\hypertarget{comparison-of-geodesic-interpolation-and-givens-interpolation}{%
\section{Comparison of geodesic interpolation and Givens
interpolation}\label{comparison-of-geodesic-interpolation-and-givens-interpolation}}

The \texttt{givens\_full\_path()} function returns the intermediate
interpolation step projections for a given number of steps. The code
chunk below demonstrates the interpolation between 2 random bases in 5
steps.

\begin{Shaded}
\begin{Highlighting}[]
\FunctionTok{givens\_full\_path}\NormalTok{(base1, base2, }\AttributeTok{nsteps =} \DecValTok{5}\NormalTok{)}
\end{Highlighting}
\end{Shaded}

\begin{figure}

{\centering \includegraphics[width=0.8\linewidth]{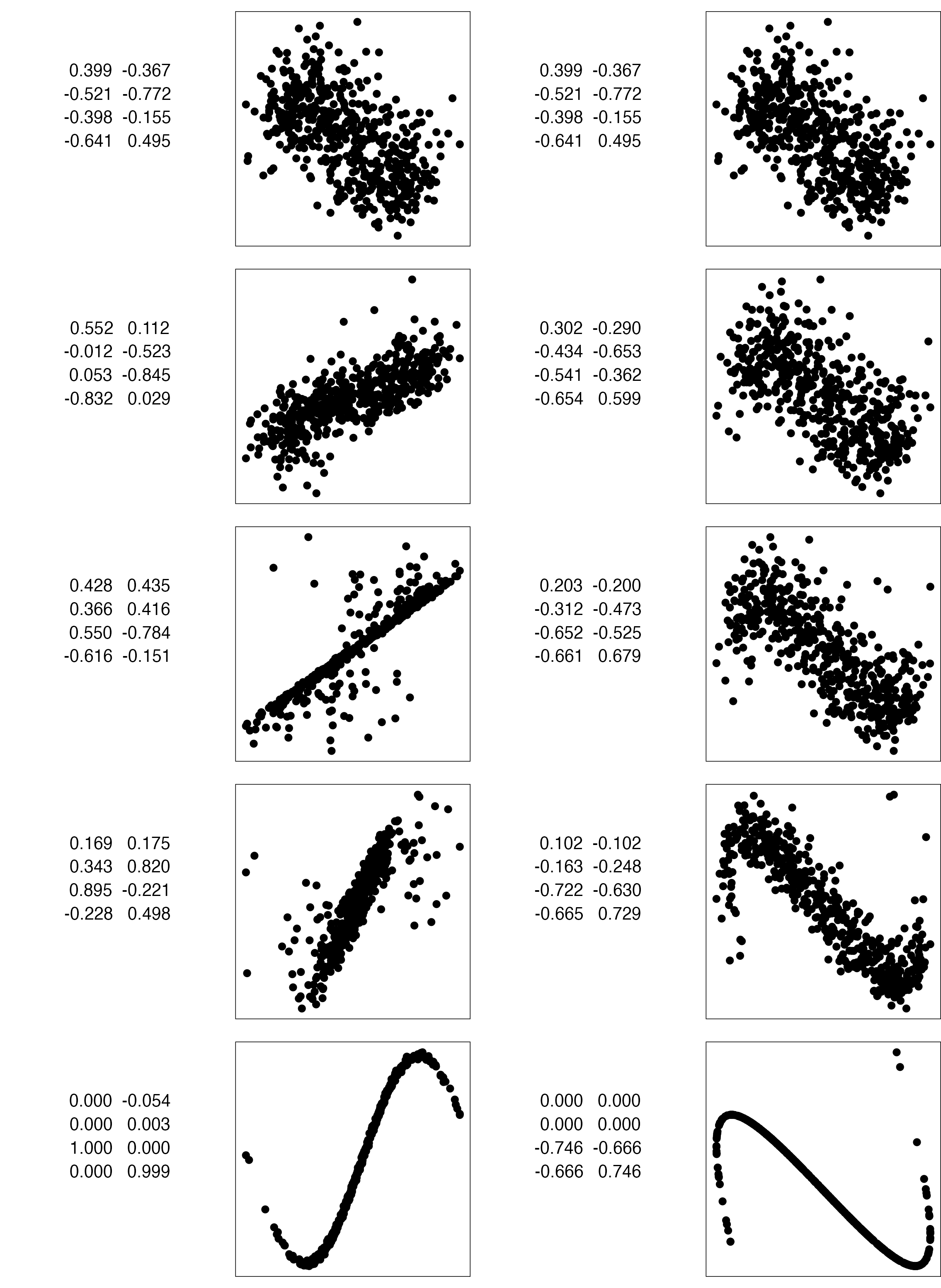} 

}

\caption{Comparison of Givens path and geodesic path between 2D projections. The Givens path preserves the frame ending at the provided basis (frame), while geodesic is agnostic to the particular basis. In general the geodesic is preferred because it removes within-plane spin, but occasionally it is helpful to very specifically arrive at the prescribed basis.}\label{fig:compare-paths}
\end{figure}

To compare the path generated with the Givens interpolation to that
found with geodesic interpolations we look at the rotation of the sine
data shown in Figure \ref{fig:splines2d-static}. We consider a subset in
\(p=4\) dimensions where the first two dimensions contain noise and the
last two contain the sine curve. Starting from a random projection we
want to interpolate towards the original sine curve. The path comparison
is shown in Figure \ref{fig:compare-paths}.

\textbf{Plotting the interpolated paths}

For further comparison and to check that the interpolation is moving in
equally sized steps we directly plot the interpolated paths. The space
of 1D projections defines a unit sphere, while 2D projections define a
torus. To illustrate the space, points on the surface of the sphere and
the torus shape are randomly generated by functions from the
\CRANpkg{geozoo} package (Schloerke 2016). The interpolated paths are
then compared within that space.

\begin{figure}

{\centering \includegraphics[width=0.45\linewidth]{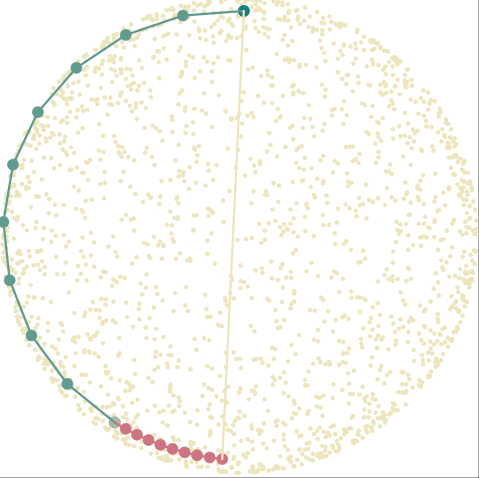} \includegraphics[width=0.45\linewidth]{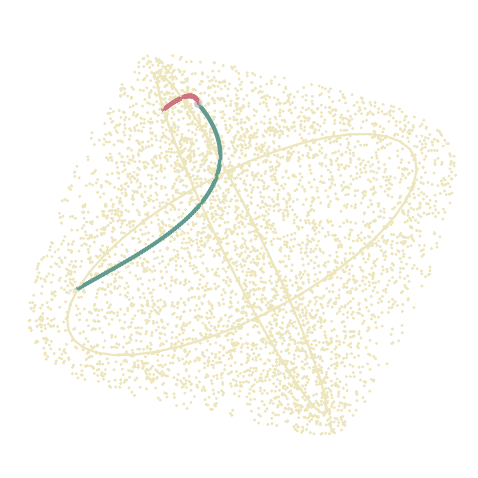} 

}

\caption{Interpolation steps of 1D (left) and 2D (right) projections of 3D data made with a Givens path (forest green) and a geodesic path (deep red). The cream points represent the space of all projections, which is a sphere for 1D projections and a torus for 2D projections. In the 1D example, geodesic arrives at the opposite side of the sphere to Givens, indicating that it has flipped the direction of the vector in order to make the shortest path to the same plane. A similar thing happens for the 2D example, geodesic flips the sign of one basis vector, but it defines the same plane, as indicated by the cream circles.}\label{fig:1d-path-static}
\end{figure}

A 1D projection of data in \(p\) dimensions corresponds to a linear
combination where the weights are normalized. Therefore, we can plot the
point on the surface of a hypersphere. In this case the Givens
interpolation will reach the exact point, while geodesic interpolation
might flip the direction and reach a point on the opposite side of the
hypersphere. Figure \ref{fig:1d-path-static} (left) shows the comparison
of the interpolation steps using the same target plane, for an example
with \(p=3\). Because the flipped target is close to the starting plane,
the geodesic path is a lot shorter.

In the case of 2D projections, we can plot the interpolated path between
2 frames on a torus. A torus can be seen as a crossing of 2 circles that
are orthonormal, as is the case with our projection onto 2D. Figure
\ref{fig:1d-path-static} (right) compares the interpolation paths for
\(p=3\).

For a 2D projection, the same target plane is found when rotating the
basis within the plane, or when reflecting across one of the two
directions (the reflected basis can then also be rotated). This means
the space of target bases is constrained to two circles on the torus,
and these are disconnected because the reflection corresponds to a jump.
In the high-dimensional space, we can imagine the reflection as flipping
over the target plane, resulting in a reflection of the normal vector on
the plan. While the Givens interpolation will reach the exact basis, a
geodesic interpolation towards the same target plane can land anywhere
along those two circles, depending on the starting basis in the
interpolation.

\hypertarget{application}{%
\section{Application}\label{application}}

This section describes the application of the Givens interpolation path
with a guided tour to explore non-linear association in multivariate
data. We use cross-rates for currencies relative to the US dollar. A
cross-rate is an exchange rate between two currencies computed by
reference to a third currency, usually the US dollar. A strong
non-linear functional relationship would indicate that underlying the
collection of cross-currency rates is a single latent factor explaining
all the movement in that time period.

\begin{figure}
\includegraphics[width=1\linewidth]{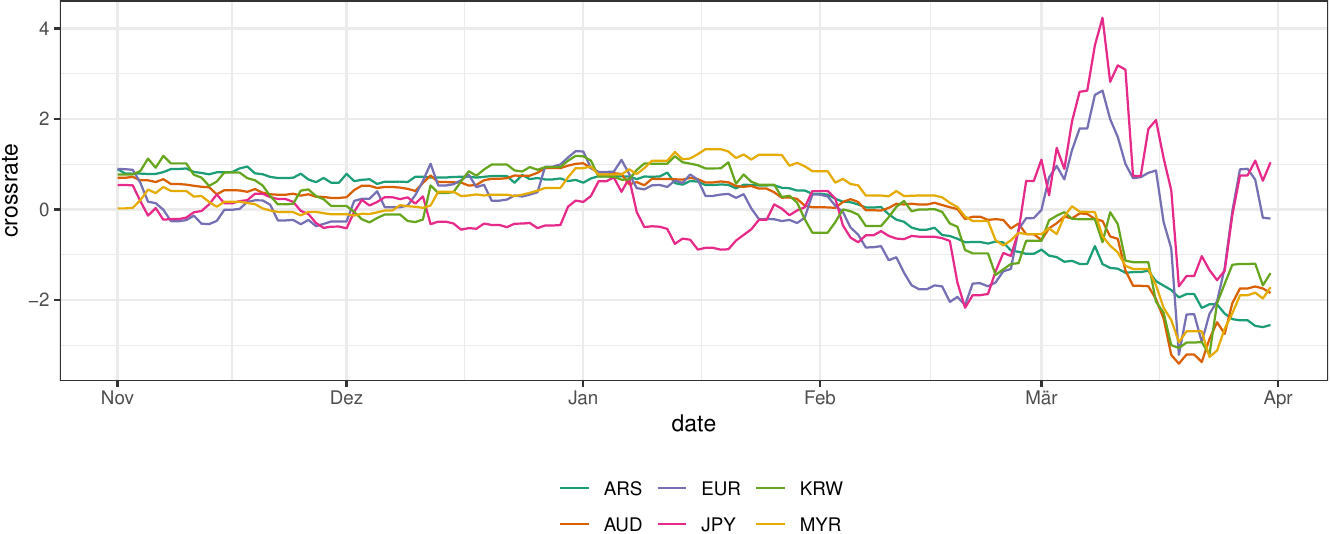} \caption{Examining the behaviour of six cross-currency rates (ARS, EUR, KRW, AUD, JPY, MYR) prior to and in the first month of the pandemic. All the currencies are standardised and the sign is flipped. JPY and EUR strengthened against the USD (high values) in March, while the other currencies weakened (low values).}\label{fig:currency}
\end{figure}

The data was extracted from \href{https://openexchangerates.org}{open
exchange rates} and contains cross-rate for ARS, AUD, EUR, JPY, KRW, MYR
between 2019-11-1 to 2020-03-31. Figure \ref{fig:currency} shows how the
currencies changed relative to USD over the time period. We see some
collective behavior in March of 2020 with EUR and JPY increasing in a
similar manner, and smaller currencies decreasing in value. This could
be understood as a consequence of flight-to-quality at these uncertain
times.

We are interested in capturing this relation over time, and from the
time series visualization we expect that we can capture the main
dynamics in a two-dimensional projection from the six-dimensional space
of currencies. Thus we start from \(p=6\) (the different currencies) and
\(n=152\) the number of days in our sample. We expect that a projection
onto \(d=2\) dimensions should capture the relation between the two
groups of currencies mentioned above, and this should be identified
within the noise of the random fluctuations.

\begin{figure}

{\centering \includegraphics[width=0.8\linewidth]{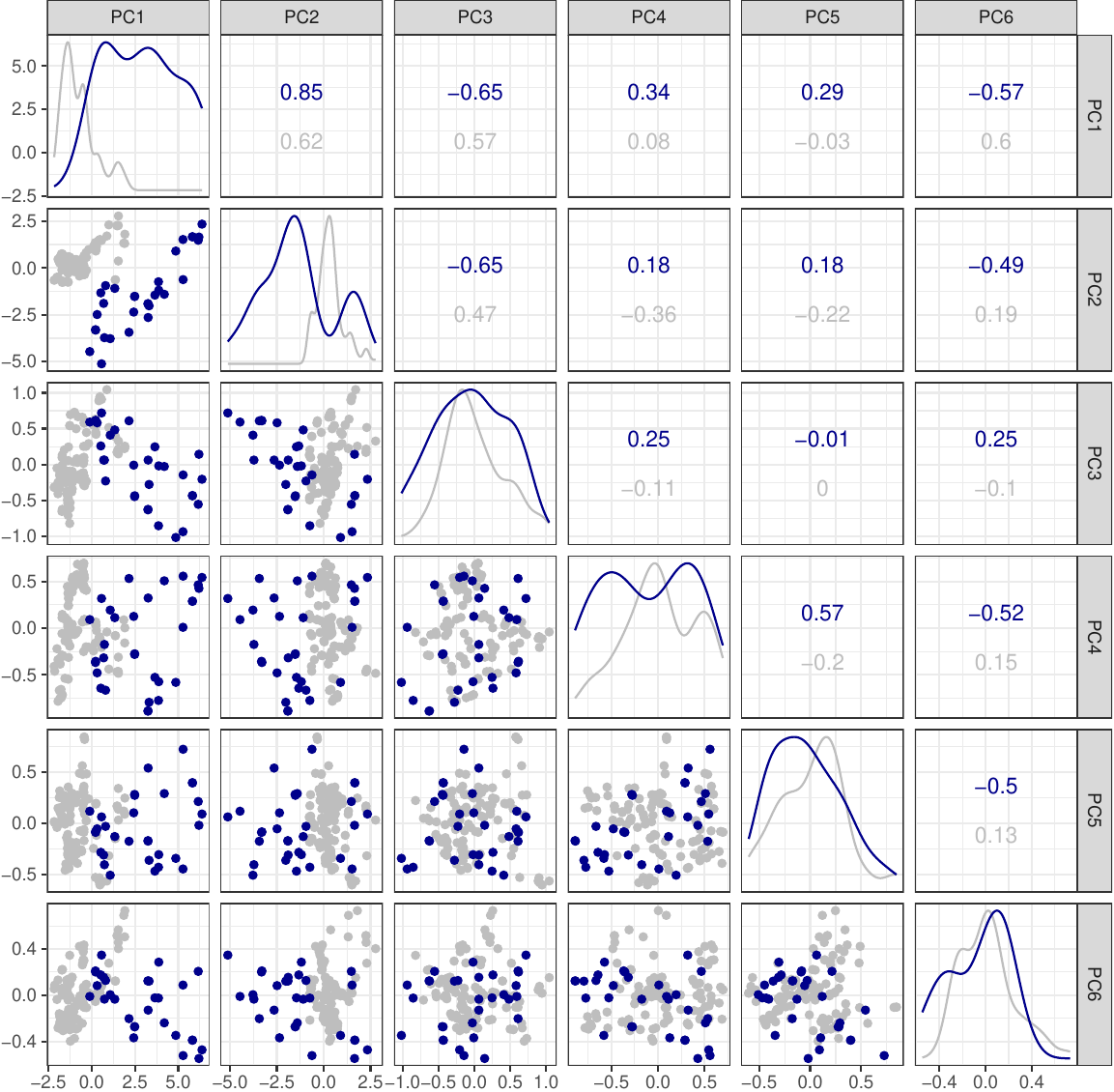} 

}

\caption{Scatterplot matrix of the six principal components. Observations in March 2020 are highlighted in dark blue, all other months are shown in grey. There is a strong non-linear dependence between PC1 and PC2.}\label{fig:pca-result-static}
\end{figure}

We may expect that we can capture the dependence between the two groups
using principal components analysis (PCA). Figure
\ref{fig:pca-result-static} shows a scatter plot matrix of the principal
components (PCs) of our dataset. Indeed we find a strong non-linear
association between the first two PCs. Investigation of the rotation
shows that the first PC is primarily a balanced combination between ARS,
AUD, KRW and MYR (and a smaller contribution of EUR), and the second
contribution is dominated by EUR and JPY contrasted with smaller
contributions from ARS and MYR. Our next step is to use projection
pursuit to further explore for non-linear functional dependence.

\textbf{Guided tour optimization}

We now use the splines index to further explore functional dependence
beyond what is observed from PCA. Note here that because of strong
linear correlations between the currencies, we start from the first four
PCs, explaining over 97\% of the variance in the data. To make the
challenge of finding functional dependence harder we start from a
projection onto the third and forth PCs. We hope to find a stronger
functional dependence than uncovered by PCA. The \texttt{display\_pca()}
function in \CRANpkg{tourr} is used to show the original variables, that
correspond to the 2D projection of the first four PCs, so that
interpretation can be made with respect to the cross-currencies.

There are several options for optimization in the \CRANpkg{tourr}
package. Geodesic optimization (GEOO) (provided by the
\texttt{search\_geodesic()} function) is a derivative-free method,
approximating a stochastic optimization algorithm. From any starting
plane, a random search in the near neighborhood delivers the most
promising direction to follow. The tour using geodesic interpolation
follows this direction until the index value decreases, and then a new
direction search is conducted. It guarantees that the index value always
increases, but it can be slow, and it can get trapped at local maxima.
Typically the best method is the simulated annealing (provided by the
\texttt{search\_better()} function) performs a random search of a large
neighborhood and then cools down the neighborhood size as the
optimization progresses. We have used this approach inserting geodesic
(SAGEO) and Givens interpolation (SAGIV) to move between targets.

\begin{figure}

{\centering \includegraphics[width=0.3\linewidth]{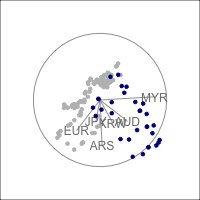} \includegraphics[width=0.3\linewidth]{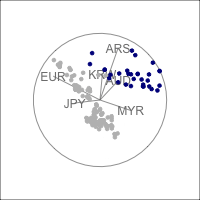} \includegraphics[width=0.3\linewidth]{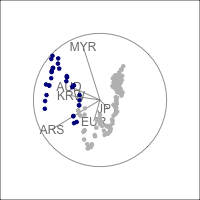} 

}

\caption{Final view after optimization in the guided tour using geodesic optimization  (GEOO) (left), simulated annealing with geodesic interpolation (SAGEO) (middle) and simulated annealing with Givens interpolation (SAGIV) (right).}\label{fig:rates-tour-static}
\end{figure}

The results are shown in Figure \ref{fig:rates-tour-static} and we see
on the right that the guided tour with SAGIV has identified a non-linear
functional dependence between the x and y axis in the final projection.
The result on the left is using GEOO, and the result in the middle uses
SAGEO. Since these methods do not allow for within-plane rotation they
were not able to identify the same and best functional relationship even
though they are using the same starting plane and index function.

\begin{figure}

{\centering \includegraphics[width=0.8\linewidth]{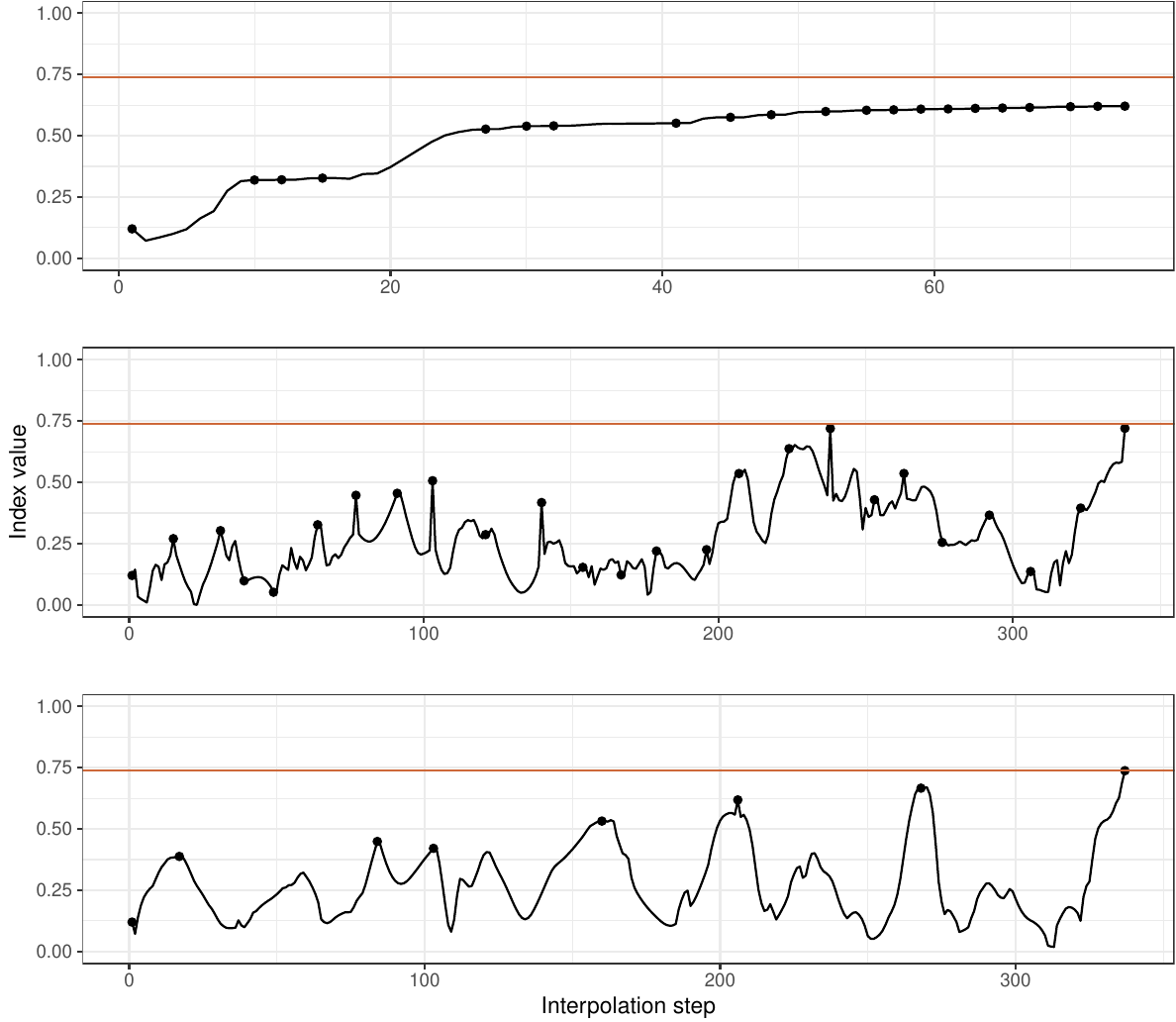} 

}

\caption{Splines index value along the interpolated optimization path in the guided tour using geodesic optimization (GEOO) (top), simulated annealing with geodesic interpolation  (SAGEO) (middle) and with Givens interpolation (SAGIV) (bottom). Points indicate index values at target planes selected during the optimization, and the horizontal line shows the maximum across the three searches. With GEOO no within-plane rotation is possible and although the optimization always heads towards higher index values it finishes at a different frame in the same plane, with a lower than possible index value. With SAGEO, a different frame in the same plane interferes with the optimization. This does not happen with SAGIV, allowing the search to find the optimal view.}\label{fig:rates-ferrn}
\end{figure}

To understand the result in more detail we use tools from
\CRANpkg{ferrn} (Zhang et al. 2021) to show how the index value is
changing over the optimization in Figure \ref{fig:rates-ferrn}. Points
indicate selected target planes, and are connected by lines showing the
index value along the interpolated path between them. For ease of
comparison, the horizontal line shown in all three panels indicates the
maximum index value found when using Givens interpolation.

The top row shows the path found via GEOO. This search strategy only
moves along geodesic paths to a target \emph{plane} and does not reach a
specific target \emph{frame}. By construction, this gives a smooth path.
However, we can see that as a consequence the optimization converges to
a plane that does not identify the pattern of interest.

In the middle we are showing the path found when using SAGEO. In this
case, the search can suggest any target frame, also allowing for any
within-plane rotation. The algorithm will compute the index value for
the suggested target frame and accept it if the index value is larger
than that for the current frame. However, once the target is accepted,
the interpolation step will reset the orientation within the plane,
potentially resulting in a frame that shows the same plane but with a
lower index value. This is why we see drops in the index value even from
one target plane to the next, and the optimization also does not
identify the pattern of interest, even though it comes close.

Finally, in the bottom row, we see the result of the random search with
Givens interpolation. While the index value can drop along an
interpolation path, the value is strictly increasing for the target
planes (up to small differences from numeric precision in the
interpolation), as is required for the optimization. Note also that the
search converged much faster (fewer target planes were selected). The
length of the interpolated paths is similar because for fixed angular
step size the Givens interpolation will take more steps to interpolate
between frames.

\hypertarget{conclusion}{%
\section{Conclusion}\label{conclusion}}

This paper describes the R package \CRANpkg{woylier} which provides an
implementation of the Givens interpolation path algorithm that can be
used as an alternative interpolation method for a tour. The algorithm
implemented in the \CRANpkg{woylier} package is as described in Buja et
al. (2005).

A prior implementation using C was available in the XGobi software
(Swayne, Cook, and Buja 1998). This new implementation was undertaken at
the request of an \CRANpkg{tourr} package user.

It is important to mention that \CRANpkg{woylier} package should be used
in conjunction with the \CRANpkg{tourr} package. Currently, it is
implemented for tours into 1D or 2D. A potential future development
would be to generalize the interpolation for more than 2D projections of
data, which could not be done here because it requires new mathematical
derivations.

For most purposes geodesic interpolation is desirable. The motivation
for frame-to-frame interpolation arises from a rotational invariance
problem of the current geodesic interpolation algorithm implemented in
\CRANpkg{tourr} package. Frame-to-frame interpolation provides users
with the tools to move to a specified frame, not the plane it defines.

Our application shows the Givens interpolation used with a projection
pursuit index which is not rotationally invariant, but ideal for
detecting bivariate non-linear association in high-dimensional data. The
potential applications where this might be useful are many and varied,
possibly facial recognition algorithms and morphing between images, or
imputing missing time steps in high-dimensional time series.

\hypertarget{acknowledgements}{%
\section*{Acknowledgements}\label{acknowledgements}}
\addcontentsline{toc}{section}{Acknowledgements}

We thank Saurabh Jhanjee for his contributions at the start of the
project. This article is created using \CRANpkg{knitr} (Xie 2022) and
\CRANpkg{rmarkdown} (Allaire et al. 2022) in R. For graphs we have used
\CRANpkg{ggplot2} (Wickham 2016), \CRANpkg{cowplot} (Wilke 2020) and
\CRANpkg{GGally} (Schloerke et al. 2021), for tables we have used
\CRANpkg{kableExtra} (Zhu 2021). The source code for reproducing this
paper can be found at:
\url{https://github.com/uschiLaa/woylier-article}.

\hypertarget{references}{%
\section*{References}\label{references}}
\addcontentsline{toc}{section}{References}

\hypertarget{refs}{}
\begin{CSLReferences}{1}{0}
\leavevmode\vadjust pre{\hypertarget{ref-rmarkdown}{}}%
Allaire, JJ, Yihui Xie, Jonathan McPherson, Javier Luraschi, Kevin
Ushey, Aron Atkins, Hadley Wickham, Joe Cheng, Winston Chang, and
Richard Iannone. 2022. \emph{Rmarkdown: Dynamic Documents for r}.
\url{https://github.com/rstudio/rmarkdown}.

\leavevmode\vadjust pre{\hypertarget{ref-asimov_1985}{}}%
Asimov, D. 1985. {``The Grand Tour: A Tool for Viewing Multidimensional
Data.''} \emph{SIAM J Sci Stat Comput 6(1)}, 128--43.

\leavevmode\vadjust pre{\hypertarget{ref-buja_cook_asimov_hurley_2005}{}}%
Buja, Andreas, Dianne Cook, Daniel Asimov, and Catherine Hurley. 2005.
{``Computational Methods for High-Dimensional Rotations in Data
Visualization.''} \emph{Handbook of Statistics}, 391--413.
\url{https://doi.org/10.1016/s0169-7161(04)24014-7}.

\leavevmode\vadjust pre{\hypertarget{ref-Buja2004TheoryOD}{}}%
Buja, Andreas, Dianne Cook, Daniel Asimov, and Catherine B. Hurley.
2004. {``Theory of Dynamic Projections in High-Dimensional Data
Visualization.''} In.
\url{http://stat.wharton.upenn.edu/~buja/PAPERS/paper-dyn-proj-math.pdf}.

\leavevmode\vadjust pre{\hypertarget{ref-cook_manual_1997}{}}%
Cook, Dianne, and Andreas Buja. 1997. {``Manual Controls for
High-Dimensional Data Projections.''} \emph{Journal of Computational and
Graphical Statistics} 6 (4): 464--80.
\url{https://doi.org/10.2307/1390747}.

\leavevmode\vadjust pre{\hypertarget{ref-grandtour1995}{}}%
Cook, Dianne, Andreas Buja, Javier Cabrera, and Catherine Hurley. 1995.
{``Grand Tour and Projection Pursuit.''} \emph{Journal of Computational
and Graphical Statistics} 4 (3): 155--72.
\url{http://www.jstor.org/stable/1390844}.

\leavevmode\vadjust pre{\hypertarget{ref-matrix_computation}{}}%
Golub, Gene H., and Charles F Van Loan. 1989. \emph{Matrix
Computations}. Johns Hopkins University Press.

\leavevmode\vadjust pre{\hypertarget{ref-Grimm2016}{}}%
Grimm, Katrin. 2016. {``Kennzahlenbasierte Grafikauswahl.''} Doctoral
thesis, Universit{ä}t Augsburg.

\leavevmode\vadjust pre{\hypertarget{ref-kruskal_1969}{}}%
Kruskal, JB. 1969. {``Toward a Practical Method Which Helps Uncover the
Structure of a Set of Observations by Finding the Line Transformation
Which Optimizes a New ``Index of Condensation.''} \emph{Statistical
Computation; New York, Academic Press}, 427--40.

\leavevmode\vadjust pre{\hypertarget{ref-mmtour}{}}%
Laa, Ursula, Alex Aumann, Dianne Cook, and German Valencia. 2023. {``New
and Simplified Manual Controls for Projection and Slice Tours, with
Application to Exploring Classification Boundaries in High
Dimensions.''} \emph{Journal of Computational and Graphical Statistics}
0 (0): 1--8. \url{https://doi.org/10.1080/10618600.2023.2206459}.

\leavevmode\vadjust pre{\hypertarget{ref-pp}{}}%
Laa, Ursula, and Dianne Cook. 2020. {``Using Tours to Visually
Investigate Properties of New Projection Pursuit Indexes with
Application to Problems in Physics.''} \emph{Comput Stat 35},
1171--1205.
https://doi.org/\url{https://doi.org/10.1007/s00180-020-00954-8}.

\leavevmode\vadjust pre{\hypertarget{ref-tourrev}{}}%
Lee, Stuart, Dianne Cook, Natalia da Silva, Ursula Laa, Nicholas
Spyrison, Earo Wang, and H. Sherry Zhang. 2022. {``The State-of-the-Art
on Tours for Dynamic Visualization of High-Dimensional Data.''}
\emph{WIREs Computational Statistics} 14 (4): e1573.
https://doi.org/\url{https://doi.org/10.1002/wics.1573}.

\leavevmode\vadjust pre{\hypertarget{ref-geozoo}{}}%
Schloerke, Barret. 2016. \emph{Geozoo: Zoo of Geometric Objects}.
\url{https://CRAN.R-project.org/package=geozoo}.

\leavevmode\vadjust pre{\hypertarget{ref-ggally}{}}%
Schloerke, Barret, Di Cook, Joseph Larmarange, Francois Briatte, Moritz
Marbach, Edwin Thoen, Amos Elberg, and Jason Crowley. 2021.
\emph{GGally: Extension to 'Ggplot2'}.
\url{https://CRAN.R-project.org/package=GGally}.

\leavevmode\vadjust pre{\hypertarget{ref-xgobi}{}}%
Swayne, Deborah F., Dianne Cook, and Andreas Buja. 1998. {``XGobi:
Interactive Dynamic Data Visualization in the x Window System.''}
\emph{Journal of Computational and Graphical Statistics} 7 (1): 113--30.
\url{https://doi.org/10.1080/10618600.1998.10474764}.

\leavevmode\vadjust pre{\hypertarget{ref-ggplot2}{}}%
Wickham, Hadley. 2016. \emph{Ggplot2: Elegant Graphics for Data
Analysis}. Springer-Verlag New York.
\url{https://ggplot2.tidyverse.org}.

\leavevmode\vadjust pre{\hypertarget{ref-tourr}{}}%
Wickham, Hadley, Dianne Cook, Heike Hofmann, and Andreas Buja. 2011.
{``{tourr}: An {R} Package for Exploring Multivariate Data with
Projections.''} \emph{Journal of Statistical Software} 40 (2): 1--18.
\url{https://doi.org/10.18637/jss.v040.i02}.

\leavevmode\vadjust pre{\hypertarget{ref-cowplot}{}}%
Wilke, Claus O. 2020. \emph{Cowplot: Streamlined Plot Theme and Plot
Annotations for 'Ggplot2'}.
\url{https://CRAN.R-project.org/package=cowplot}.

\leavevmode\vadjust pre{\hypertarget{ref-mgcv-gam}{}}%
Wood, S. N. 2011. {``Fast Stable Restricted Maximum Likelihood and
Marginal Likelihood Estimation of Semiparametric Generalized Linear
Models.''} \emph{Journal of the Royal Statistical Society (B)} 73 (1):
3--36.

\leavevmode\vadjust pre{\hypertarget{ref-knitr}{}}%
Xie, Yihui. 2022. \emph{Knitr: A General-Purpose Package for Dynamic
Report Generation in r}. \url{https://yihui.org/knitr/}.

\leavevmode\vadjust pre{\hypertarget{ref-ferrn}{}}%
Zhang, H. Sherry, Dianne Cook, Ursula Laa, Nicolas Langrené, and
Patricia Menéndez. 2021. {``Visual Diagnostics for Constrained
Optimisation with Application to Guided Tours.''} \emph{The R Journal}
13 (2): 624--41. \url{https://doi.org/10.32614/RJ-2021-105}.

\leavevmode\vadjust pre{\hypertarget{ref-kableextra}{}}%
Zhu, Hao. 2021. \emph{kableExtra: Construct Complex Table with 'Kable'
and Pipe Syntax}. \url{https://CRAN.R-project.org/package=kableExtra}.

\end{CSLReferences}

\bibliographystyle{unsrt}
\bibliography{woylier-article.bib}

\end{document}